\documentclass{aa}
\usepackage{epsfig}

\def\iue{\mbox{\it IUE}}
\def\ines{\mbox{\it INES}}
\def\swet{\mbox{\it SWET}}
\def\newsips{\mbox{\it NEWSIPS}}
\def\iuesips{\mbox{\it IUESIPS}}

\def\nus{\mbox{$\nu$}}

\def\back2{\mbox{$B(x,\lambda)$}}

\def\sig2{\mbox{$\sigma(x,\lambda)$}}
 
\begin{document}

\thesaurus{13(03.13.2; 03.19.2; 03.20.1; 13.21.2)}


\title{Improved data extraction procedures for IUE Low Resolution Spectra: The 
INES System}

\author{P.M. Rodr\'{\i}guez-Pascual\inst{1} \fnmsep 
\thanks{{\it Previously}: ESA - IUE Observatory}
\and
R. Gonz\'alez-Riestra\inst{2} \fnmsep$^{\star}$
\and N. Schartel\inst{3}\fnmsep$^{\star}$\fnmsep
\thanks{Affiliated to the Astrophysics Division, SSD, ESTEC} 
\and W. Wamsteker\inst{4}\fnmsep $^{\star\star}$}

\institute{
Departamento de F\'{\i}sica, Universidad Europea de Madrid,  C/Tajo S/N, 28670 
Villaviciosa de Od\'on, Spain
\and 
LAEFF, P.O. Box 50727, 28080 Madrid, Spain
\and 
ESA - XMM SOC, P.O. Box 50727, 28080 Madrid, Spain
\and ESA - IUE Observatory , P.O. Box 50727, 28080 Madrid, Spain}

\date{Received / Accepted}

\titlerunning{INES Low resolution data}
\authorrunning{Rodr\'{\i}guez-Pascual et al.}

\offprints{P.M. Rodr\'{\i}guez-Pascual}
\mail{p\_miguel.rodriguez@fis.cie.uem.es}

\maketitle

\begin{abstract}

We present the extraction and processing of the \iue\ Low Dispersion
spectra within the framework of the ESA ``IUE Newly Extracted Spectra''
(\ines) System. Weak points of \swet, the optimal extraction 
implementation to produce the \newsips\ output products (extracted spectra)
are discussed, and the procedures implemented in
\ines\ to solve these problems are outlined.  The more relevant
modifications are: 1) the use of a new noise model, 2) a more accurate
representation of the spatial profile of the spectrum and 3) a more
reliable determination of the background. The \ines\ extraction also
includes a correction for the contamination by solar light in long
wavelength spectra. Examples showing the improvements obtained in \ines\
with respect to \swet\ are described. Finally, the linearity and
repeatability characteristics of \ines\ data are evaluated and the validity
of the errors provided in the extraction is discussed.


\end{abstract}

\section{Introduction}

The International Ultraviolet Explorer (\iue) collected more than 104000
spectra of all types of astronomical objects during its more than 18 years
of operations. The \iue\ Project considered it desirable to make available
to the astronomical community a ``Final Archive'' holding all the \iue\
data processed in an uniform way and with improved reduction techniques and
calibrations.  For this purpose a new processing system (\newsips) was
developed and the full \iue\ archive was re-processed with a newly derived
linearization and wavelength scale. Also an adapted optimal extraction
scheme (Horne 1986), \swet, was used to derive the low resolution
absolutely calibrated output spectra. 
A full description of the \newsips\ system is given in Nichols \& Linsky
(1996) and in Nichols (1998). Technical details can be found in the
\newsips\ Manual (Garhart et al. 1997).

One of the main goals of the system is to obtain the maximum
signal-to-noise ratio in the final data.  For this purpose the geometric
and photometric corrections are performed through a new approach, based on
cross-correlation techniques to align science and Intensity Transfer
Functions (ITF) images (Linde \& Dravins 1990). The application of this new
approach reduces substantially the fixed pattern noise, and leads to
improvements in the signal-to-noise ratio between 50 and 100\%\ in low
dispersion spectra and between 50 and 200\% in high resolution data
(Nichols 1998).

The intrinsic non-linearity of the detectors (SEC VIDICON cameras) makes
the photometric correction one of the most critical tasks in the processing
of \iue\ data. The correction is performed through the Intensity Transfer
Functions (ITFs), which are derived from series of graded lamp
exposures. These functions transform the raw Data Numbers (DN) of each
pixel in the Raw Image into linearized Flux Numbers (FN) in the
Photometrically Corrected Image. Specifically for the Final Archive, a new
set of ITFs images were obtained for the three cameras under well
controlled spacecraft conditions and through improved algorithms.  However,
the final extracted spectra still show some residual non-linearities, most
likely due to the breakdown at the extreme ITF levels of the assumption
that over small differential flux ranges the relation between FNs and DNs
can be approximated by a linear interpolation.

Further modifications implemented in \newsips\ include the improvement in
the wavelength calibration, the revision of the flux scale, the derivation
of noise models and the optimal extraction of spectra (only for low
resolution). The existence of noise models has allowed to estimate the
errors on \iue\ fluxes for the first time. A special effort has also been
made to ensure the correctness of all the information referring to the
specific observation attached to the data.

The quality control procedures applied by the IUE Project have shown that
the \newsips\ reprocessed spectra are superior to the \iuesips\ spectra in
all cases (Nichols 1998). For the high resolution spectra the new methods
to estimate the image background (Smith 1998) and the ripple correction
algorithm (Cassatella et al. 1998) result in a much higher quality high
resolution spectra for this data. However, it was found that the low
resolution data extraction still contained some serious shortcomings which
would affect significantly the usefulness of the extracted spectra.
(Talavera et al. 1992, Nichols 1998).  Most of these shortcomings and
drawbacks in the IUEFA products were related to the method for the final
extraction of the 1-D spectra (\swet) from the bi-dimensional, spatially
resolved, rotated images (SILO\footnote{The photometrically and
geometrically corrected image is rotated so that the dispersion direction
is along the X axis.}) files.

Within the framework of the ESA IUE Data Distribution System, it was
decided to correct all the low dispersion spectra through the application
of new extraction algorithms that significantly improve the quality and
reliability in the final data products.  A completely different philosophy
is behind these new algorithms. The model-dependent strategy followed in
\swet\ is abandoned, with the aim of retaining as much information as
possible concerning the data. We anticipate that the results of both
techniques are essentially identical, when the model parameters used by
\swet\ are well suited, namely, for well exposed continuum sources. The
method chosen has assured that the improvements achieved with the \newsips\
geometric and photometric corrections are preserved since the new
algorithms work on the SILO files. In this paper we describe the main
features of the \ines\ extraction procedures: background and spatial
profile determination, quality flags handling, solar contamination removal,
homogenization of the wavelength scale (Section 2).

In Section 3 the repeatability, errors reliability and 
linearity of \ines\ low dispersion data are evaluated. Finally, the major
improvements achieved by \ines\ are summarized in Section 4.

\section{Optimal extraction of IUE Low dispersion spectra}
	\label{sec:introlo}

Optimal extraction techniques for bidimensional detectors were originally 
developed for CCD chips (Horne 1986). The basic equations of the method are:
\begin{equation}  
FN(\lambda) =  
\frac{ \sum_{x}{}  
[FN(x,\lambda)-B(x,\lambda)]   
\frac{p(x,\lambda)}{\sigma(x,\lambda)^2}}  
{\sum_{x}{ }\frac{p(x,\lambda)^{2}}{\sigma(x,\lambda)^2}}  
\label{eq:optimal} 
\end{equation}  
\begin{equation}  
\frac{1}{\Delta FN(\lambda)^{2}}={\sum_{x}{ }  
\frac{p(x,\lambda)^2}{\sigma(x,\lambda)^2}}  
\end{equation}  
where the variables are:  
\begin{itemize}  
\item{} $x$\ : coordinate in the cross-dispersion (spatial) direction  
\item{} $\lambda$\ : coordinate in the spectral direction  
\item{} $FN(x,\lambda)$\ : FN value at pixel $(x,\lambda)$  
\item{} $B(x,\lambda)$\ : background at pixel $(x,\lambda)$  
\item{} $\sigma(x,\lambda)$\ : noise at pixel $(x,\lambda)$  
\item{} $p(x,\lambda)$\ : extraction profile at pixel $(x,\lambda)$  
\item{} $FN(\lambda)$\ : total flux number (FN) at $\lambda$   
\end{itemize}  
 
It must be noted that the \iue\ detectors are quite different from CCDs,
which are nearly linear detectors, with a very large dynamic range, formed
by individual pixels, almost independent on their neighbors. None of these
characteristics are valid for the \iue\ SEC Vidicon cameras. Each raw \iue\
image consist on a 768x768 array of 8-bit elements, which are not physical
pixels, but picture elements determined by the stepping and size of the
camera readout beam. The focusing system of this beam introduces geometric
distortions. The dynamic range is small (0-255 DN) and the response of the
camera is non-uniform and highly non-linear. Furthermore, the noise in
these detectors deviates strongly from the poissonian photon noise of CCDs.

Therefore, a direct application of the techniques used for CCDs is not
appropriate. The application of Eq. 1 to \iue\ data requires a careful
determination of the noise model, the background estimation, the extraction
profile and the treatment of ``bad'' pixels. Furthermore, these
determinations are model dependent and the best results are obtained only
after fine tuning a number of parameters. In an interactive processing the
choice of the best set of parameters is made case by case. For automatic
processing, these processing parameters are fixed and must be chosen so as
to cover the largest number of possible cases. This approach unavoidably
leads to the degradation of the performance of the system.

In the following subsections, the main items entering in the optimal
extraction process are discussed, indicating the solutions adopted in
\swet\ and identifying the problems which have led to the different
extraction scheme applied in \ines.

\subsection{Noise models}
	\label{sec:noise}

The estimate of the noise in \iue\ data is essential at two stages of the 
extraction of the spectrum from the SILO images.  Firstly, the determination 
of the extraction profile requires an evaluation of the signal-to-noise 
ratio in order to perform weighted fits to the data.  Secondly, the errors in 
individual pixels are propagated through the extraction procedure in order 
to assign errors to the final extracted fluxes. 

\begin{figure}[htb] 
\begin{center} 
\psfig{figure=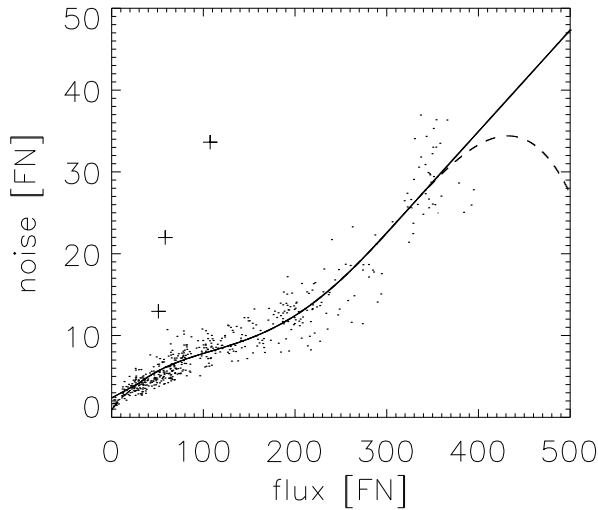,width=9.cm 
       ,bbllx=2.5cm,bblly=13.8cm,bburx=13.0cm,bbury=22.0cm,clip=} 
\caption{Example of the noise model for the wavelength 1612\AA . Points
represent the standard deviation $\sigma({\rm FN})$\ as a function of the
corresponding median flux numbers $<{\rm FN}>_{0}$. The crosses indicate
data not considered for the modeling.  The continuous line shows the noise
model applied in \ines\ consisting of a fitted polynomial and its linear
extrapolation.  The broken line shows the \swet\ extrapolated polynomial
for comparison. }
\label{fig:errex} 
\end{center} 
\end{figure}

The characteristics of the noise in the \iue\ Raw images are strongly
altered by the photometric linearization procedure via the Intensity
Transfer Function and by the spatial resampling required to derive the SILO
image format. The approach followed to derive the noise model in \swet, as
well as in \ines, has been to model it empirically from SILO science and
lamp (UV-flood) images (Garhart et al. 1997). However the final noise
models in the \ines\ procedure are different from that used in \swet\ in
two points: the extrapolation to high FN values and the handling of very
low FN values. In the first case, the \swet\ noise model extrapolates a
third order polynomials determined from the fitting of lower FN's. These
polynomials often have a negative derivative for high FN values, leading to
unrealistic estimates of the noise as shown in Figure~\ref{fig:errex}. At
the low end of the FN range, \swet\ noise models also use high order
polynomials which introduces strong boundary effects. It is especially
remarkable that \swet\ assigns an error of 1FN to negative FNs, which occurs
because of statistical fluctuations around the adopted NULL ITF level

In the \ines\ noise models, for every wavelength interval the standard
deviation as a function of the FN is described by polynomials of different
order for different FN ranges.  For FN values below thirty, a first order
polynomial was used in order to avoid boundary effects.  In the FN range
from thirty up to the point where still enough data points are available
(``breakpoint'') a higher order polynomial was used (third degree for LWP,
fourth for SWP and LWR). The region of higher FN is linearly extrapolated
based on the third (fourth) order polynomial fit
(Fig.~\ref{fig:errex}). Therefore, for a given wavelength, $\lambda$ the
noise, $\sigma$(FN), is represented by:
 
\begin{eqnarray*} 
\lefteqn{\sigma(FN)|_{\lambda=const.} =} \\ 
& \left\{  
      \begin{array}{lll} 
      B_1 + C_1 \cdot FN                         		& \;\; for \;\;&  FN 
\le  30  \\ \\ 
\sum\limits_{i=0}^{4_{(SWP)},3_{(LWP)} } A_{i} \cdot  FN^{i} 	& \;\; for 
\;\;&  30 \le FN \le breakpoint \\ \\ 
      B_2 + C_2 \cdot FN                          		& \;\; for 
\;\;&  FN > breakpoint  
      \end{array} \right. 
\end{eqnarray*} 
 
The fitting of the third (fourth) order polynomial was iterated five times, 
excluding data points for which $\sigma$(FN)was greater then two times the 
values fitted in the previous iteration in order to exclude cosmic rays 
and similar features. 
 
The extrapolation to high FN values was based on the fifty highest data
points. The ``breakpoint'' was defined as the value with the largest
positive derivative. For the LWP camera the ``breakpoints'' are found at
values between 390 and 460 FN, depending on the wavelength. For SWP they
are in the range 280 to 400 FN, and for LWR in the range 105 to
410. Therefore the extrapolation in the LWR camera covers a larger range of
FNs.
 
The noise models were smoothed in the wavelength direction following a 
similar approach, i.e. different polynomials were used for different 
cameras and wavelength ranges. 
 
Finally, the noise model was interpolated over a two dimensional grid of 
1025 FN values (from 0 to 1024) by 640 pixels in the wavelength direction. 
In the cases in which the SILO file has negative FN values, the noise of 
these pixels is taken as the value corresponding to FN=0 for that 
wavelength.

As expected, both noise models are indistinguishable for most FN values and
wavelengths. It is only in very short exposure time images and/or images
with pixels reaching FN values larger than the "break-points" defined above
that different results are obtained. It should be noticed that in the
\swet\ method a single high FN pixel with an incorrectly extrapolated error may affect
significantly the extraction profile determination because of the
exceptional signal-to-noise ratio assigned to it.

\subsection{Spectral Extraction}
	\label{sec:extrac}

According to Eq.~1, the three major items in the optimal extraction method
are: the background, the spatial profile and the noise model. Their
treatment in the \ines\ extraction procedure is described in following
subsections. In  addition, a subsection is devoted to describe the handling
of those pixels whose 
quality is non-optimal. The method applied to remove the solar
contamination in LWP images is also described in detail in a separate
subsection. Finally, the method to homogenize the wavelength scale for all
long and short wavelength spectra, independently of observing epoch or ITF,
is outlined.

    	\subsubsection{Background determination}
	\label{sec:bkg}

The background in \iue\ science images is a combination of different
sources: particle radiation, radioactive phosphor decay in the detector,
halation within the UV converter, background skylight, scattered light and
readout noise. The first two depend on the instrument itself and on the
radiation environment and vary slowly across the camera faceplate, whereas
the last three depend on the spectral flux distribution of the object
observed and their integrated effect varies in a complicated way across the
raw image.

The background is derived from two swathes seven pixels
wide in the spatial direction, symmetrically located with respect to the
center of the aperture. 

\begin{figure}[ht] 
\begin{center}
\psfig{file=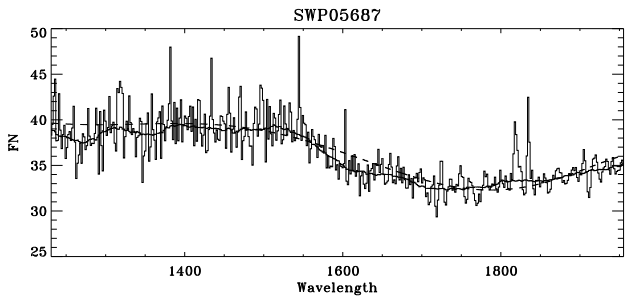,width=9.cm}
\caption{Example of background smoothing in \ines\ (solid line). The 
dashed line represents the 6th degree polynomial fit used in \swet.}
\label{fig:bkgspec}
\end{center}
\end{figure}

Along the dispersion direction, the method to estimate the background has
to remove the high frequency noise but preserving the low frequency
intrinsic variations. The two approaches generally followed in the past
have been (a) to apply consecutively a median and a box filter (\iuesips)
or (b) to fit the background to a polynomial (\swet). A direct smoothing is
simple, robust and model independent, but sensitive to bright spots and
outlying pixels. A polynomial fit is more efficient in removing such
outliers, but the degree of the polynomial must be too high to reproduce
the small scale variations. As a compromise providing acceptable solutions,
we have adopted for \ines\ an iterative method in which the background is
median and box filtered (31 pixels wide), allowing for outlying pixels
rejection in each iteration. This method effectively reduces the noise,
preserves the intrinsic background variations in relatively small scales
and removes bright spots (Fig.~\ref{fig:bkgspec}).

\begin{figure}[ht] 
\begin{center}
\psfig{file=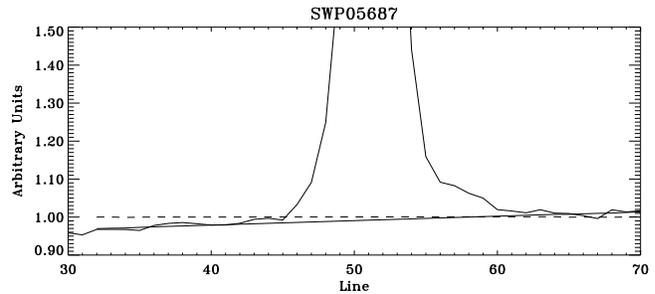,width=9.cm}
\caption{This figure shows an example where there are small, 
but significant, variations of the background in the spatial direction. The
dashed line is the average of the \swet\ background and the thick line is the
background as computed by \ines}
\label{fig:bkgprof} 
\end{center}
\end{figure} 

\iuesips\ and \swet\ assumed a constant background in the spatial
direction. For non-optimal extraction this may be an acceptable approach
since the overestimate at one side of the aperture is roughly compensated
by the underestimate at the other side. However, such compensation does not
occur in an optimization technique such as \swet\ because the weighting
profile will be forced to zero in the region where the background is
overestimated, leading to a distortion of the extraction profile with
respect to the true spatial profile (Fig.~\ref{fig:bkgprof}). In the
extraction of the \ines\ data, the background for each line within the
aperture region is obtained from a linear interpolation between the
smoothed background at both sides of the aperture.

The largest deviations between \ines\ and \swet\ results due to different
background estimates are expected in images where the net signal from the
target is rather weak. As will be described in next subsections, both
methods follow completely different approaches to obtain the final 1-D
spectrum from underexposed images. Since it is not easy to show the sole
effect of the background, we defer to next subsections the discussion of
the differences between both methods in underexposed spectra. 

    	\subsubsection{Extraction profile}
	\label{sec:prof}

The {\it not interactive} processing of the data implies that the
extraction parameters cannot be fine tuned for each individual
spectrum. Furthermore, the targets observed with \iue\ span a wide range of
properties: pure continuum/line emission, very blue/red objects,
extended/point-like sources, multiple sources within the aperture, etc.

In \ines\ the spatial profile is modeled so that it is smooth, but able to
track short scale variations along the dispersion direction. The 2-D
spectrum is blocked in bins of similar total S/N and interpolated linearly
in wavelength.  The process is iterative and outlying pixels are rejected
after each iteration.  The iteration stops when no further outliers are
found. All pixels with no real flux information (not photometrically
corrected, telemetry dropout, reseaux, permanent artifacts, 159DN corrupted
pixels) are excluded from the process of flux extraction. This method
provides results in agreement with \swet\ within 2- 3\%\ for 
{\em well exposed continuum} spectra, corresponding to the repeatability
errors of the \iue\ instruments (see Section \ref{sec:dataev}).


For very underexposed spectra where the total S/N is too low to determine
the spatial (weighting) profile empirically, the adopted approach in \ines\
is to add-up all the spectral lines within the aperture (boxcar
extraction). In contrast, the \swet\ method depends on the expected
extension of the source: for extended sources a boxcar extraction is used
too, but for point sources a default point-like extraction profile is used
{\em at} the center of the aperture.  These two different approaches define
the difference in the philosophy underneath \swet\ and \ines : \swet\ goal
is to get the highest signal-to-noise spectrum, even if at some particular
cases (weak sources that are not point-like in spite of its classification
or point-like weak sources miscentered in the aperture) the flux reported
is not correctly computed.
\ines\ goal is to get the best representation of the actual flux
at all wavelengths, even at the cost of not reaching the highest
signal-to-noise (weak point-like sources).

\begin{figure}
\begin{center}
\epsfig{file=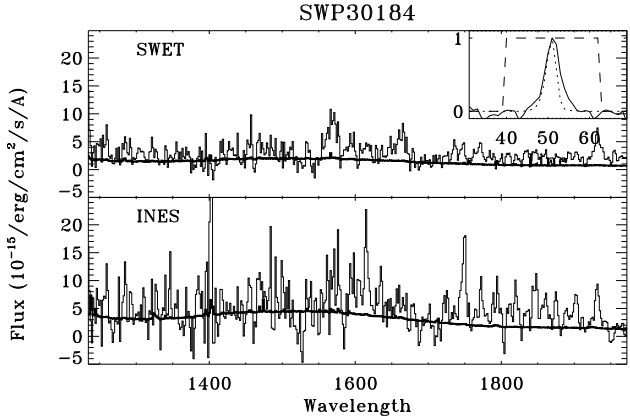,width=9.cm}
\caption{Example of the differences in the extraction of a weak point
source properly centered in the aperture. The flux level is similar with
both extractions, but the \ines\ spectrum is noisier because it has not
been optimally extracted (see text for details).
Solid lines in both panels indicate the extraction errors. The
actual spatial profile is shown in the upper right corner of the figure
together with the default profile used by \swet\ ({\em dotted line}) and
the uniform weight used by \ines\ ({\em dashed line}).}
\label{fig:weakcnt}
\end{center}
\end{figure}

\begin{figure}
\begin{center}
\epsfig{file=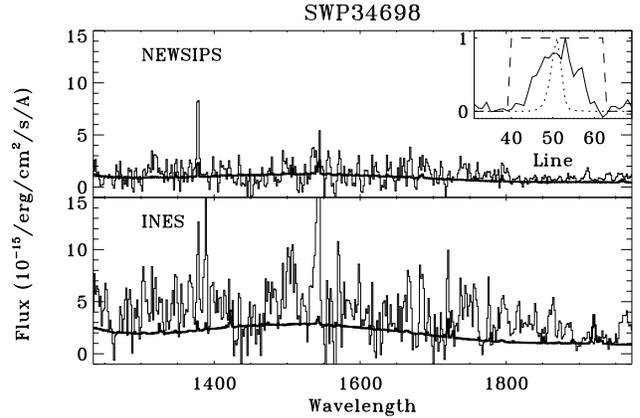,width=9.cm}
\caption{Example of the differences in the extraction of a weak extended
source. The use of a boxcar extraction in the whole aperture for weak
sources guarantees that all the flux will go into the extracted
spectrum. Symbols are as in \ref{fig:weakcnt}.}
\label{fig:weakxtd}
\end{center}
\end{figure}

\begin{figure}
\begin{center}
\epsfig{file=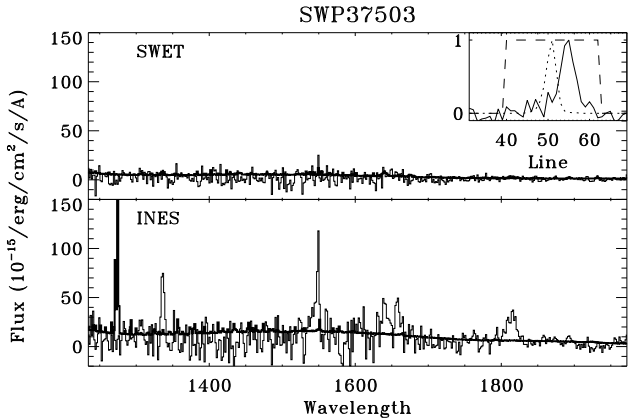,width=9.cm}
\caption{Example of the extraction of a weak point source
miscentered in the aperture.  The default extraction profile used by 
\swet\ at the expected location of the spectrum gives the largest weights to
regions where there is no spectrum, leading to an underestimate of the
flux. Symbols are as in Fig. \ref{fig:weakcnt}}
\label{fig:weakpnt}
\end{center}
\end{figure}

Figures \ref{fig:weakcnt}, \ref{fig:weakxtd}  and
\ref{fig:weakpnt} show examples of the different results obtained with
\swet\ and \ines\ for weak spectra. In all the examples, the spectrum is 
too weak for its profile to be determined empirically and the sources are
classified as point-like. Therefore, \swet\ uses a default point-source
profile and \ines\ uses a boxcar through the whole aperture. When there is
a true point source, properly centered in the aperture
(Fig.~\ref{fig:weakcnt}), both extractions provide similar flux levels,
although the \ines\ spectrum is noisier. The second example is an exposure
on the echo of the SN1987A in the Large Magellanic Cloud through the large
aperture. Since the image is classified as IUECLASS 56 (Supernova), \swet\
uses the default point-like extraction profile at the center of the
aperture (dotted line in the inset in Fig.~\ref{fig:weakxtd}), and the resulting flux is
underestimated by more than a factor 2. Obviously, the boxcar method used
by \ines\ results in a noisier spectrum, but provides the correct flux
level, better representing the actual information content of the spectrum.

SWP 37503 is an image of CC~Eri, a rapidly rotating late type star with
strong chromospheric emission lines.  Here, \swet\ again uses the default
point-like extraction profile at the aperture center
(Fig.~\ref{fig:weakpnt}).  The source is indeed point-like, but was not
properly centered within the slit. Thus, the extraction profile used by
\swet\ is offset with respect to the location of the spectrum, resulting in
a formal non-detection of the source, in particular of the strong emission
lines. The boxcar method used in \ines\ produces a noisier spectrum,
but the emission lines are correctly extracted.

Strong narrow emission lines onto a weak continuum have been reported to be
incorrectly extracted by \swet, even though they are optimally exposed
(Talavera et al. 1992, Hu\'elamo et al. 1999). The problem is that in these
cases there exist 
variations in the spatial profile on wavelength scales much shorter than
\swet\ can follow. The origin is the "beam pulling" effect (Boggess et
al. 1978) which consists in a deflection of the readout beam in regions
with large charge variations in the image section of the cameras. The shift
in the image registration can be as much as 2 lines along the
cross-dispersion direction in a few wavelength steps. The result is that
the emission line registration is shifted with respect to the continuum. If
the extraction profile cannot change on wavelength scales of the order of
the spectral resolution, the strong unresolved emission lines are
recognized and flagged as "cosmic rays", resulting in a strong
underestimate of the lines flux. To account for this effect, the \ines\
extraction method sets the minimum block size to 7 wavelength bins,
slightly larger than the spectral resolution.

\begin{figure}
\begin{center}
\psfig{file=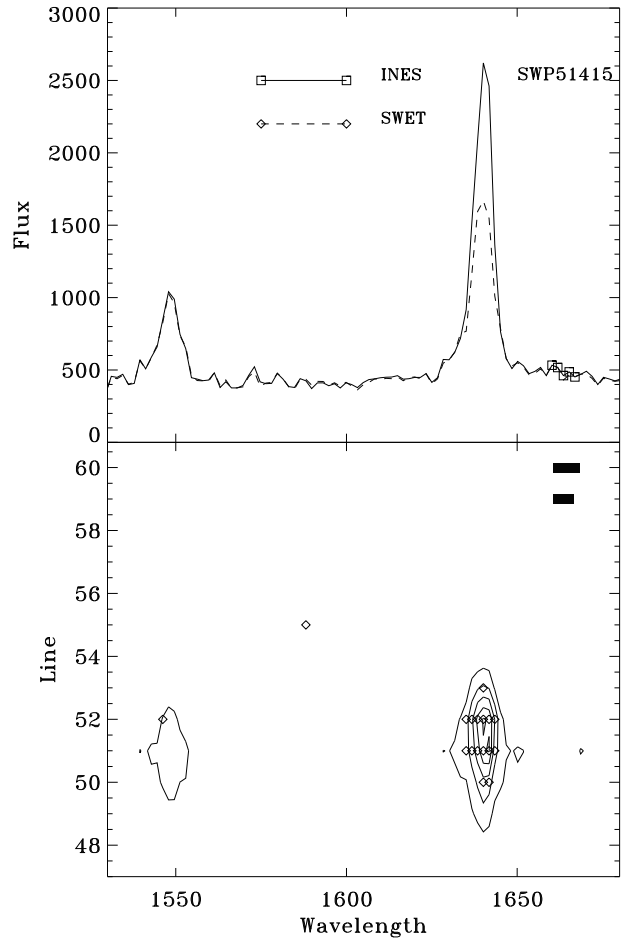,width=9.cm}
\caption{Example of the extraction with \ines\ and \swet\ of a spectrum 
with weak continuum and strong narrow emission lines in. The HeII 1640\AA\
line is not flagged in either of the extracted spectra.  Shown in the
bottom panel is the bi-dimensional SILO file with the 2-D quality flags
overplotted. Diamonds correspond to flag ``-32'' (\swet\ Cosmic ray) and
filled squares to other flags (e.g. reseau marks).}
\label{fig:fl1}
\end{center}
\end{figure}

An example of this effect is shown in Figure \ref{fig:fl1}. The
spectrum belongs to the symbiotic star AG Dra, characterized by a weak
continuum with strong narrow emission lines. The intensity of the HeII
1640\AA\ line given by \swet\ is approximately half the intensity given by
\ines . \swet\ finds part of the emission line outside the extraction profile,
consequently flags the pixels as ``cosmic rays'' (flag -32) and rejects them
in the derivation of the final spectrum. It is also worth to note that although
half the line is rejected as "cosmic ray" the flags do not go into the
final 1-D quality flag spectrum (see next subsection).

\begin{figure}
\begin{center}
\psfig{file=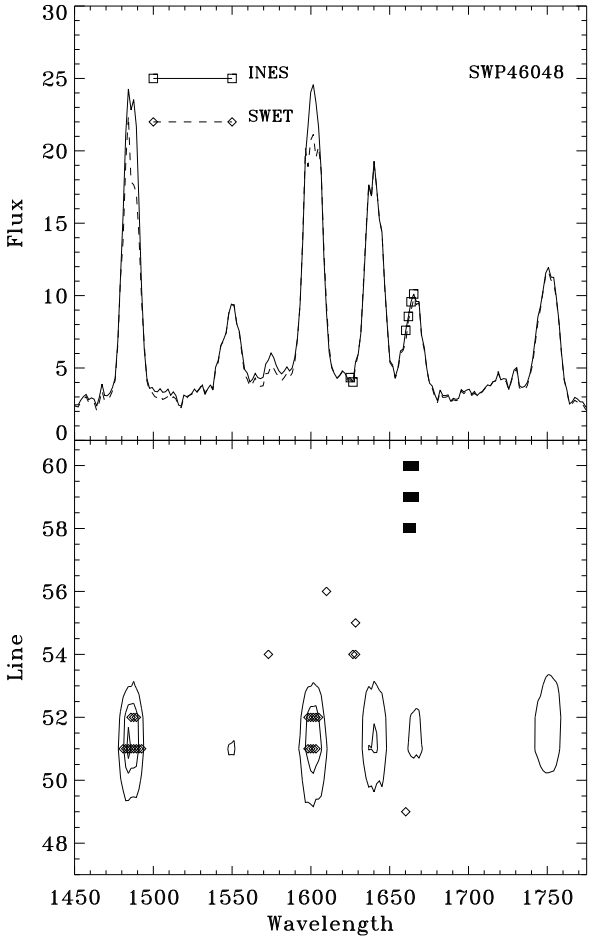,width=9.cm}
\caption{In this spectrum of Nova V351 Puppis 1991 none of the emission 
lines are flagged in either of the extractions, but the NIV] 1486\AA\ and
the [NeIV] l602\AA\ lines have been identified by \swet\ as ``cosmic
rays'', and their intensity is underestimated. Symbols are as in Figure
\ref{fig:fl1}.}
\label{fig:fl3}
\end{center}
\end{figure}

A similar example (a spectrum of Nova Puppis 1991) is shown in Figure
\ref{fig:fl3}.  The ratio NIV]1486~\AA/NIII]1750\AA\ is  
smaller by a 20\%\ when derived from the \swet\ spectrum, and clearly in
error. These examples demonstrate that \swet\ results for sources with
strong narrow emission lines onto a weak continuum are not optimal and
the use of line ratios as diagnostics for physical parameters
(temperature, density, chemical abundances\ ...) may be misleading, greatly
diminishing the usefulness of the IUEFA for general usage.

 	\subsubsection{Quality flags handling and propagation}
	\label{sec:flags}

Quality flags (\nus's) mark those SILO pixels whose quality is not
optimal. The quality of a pixel can be affected by different problems, and
there is a gradation in the reliability of the value.  Flags are coded in
\newsips\ in such a way that more negative values indicate more important
problem conditions.

The importance of a proper handling of \nus's is twofold: firstly, the
flags are used to exclude "bad" pixels during the extraction procedure and
secondly, they mark in the final 1-D extracted spectrum those wavelengths
where the user should be warned about the reliability of the flux.

\begin{figure}
\begin{center}
\psfig{file=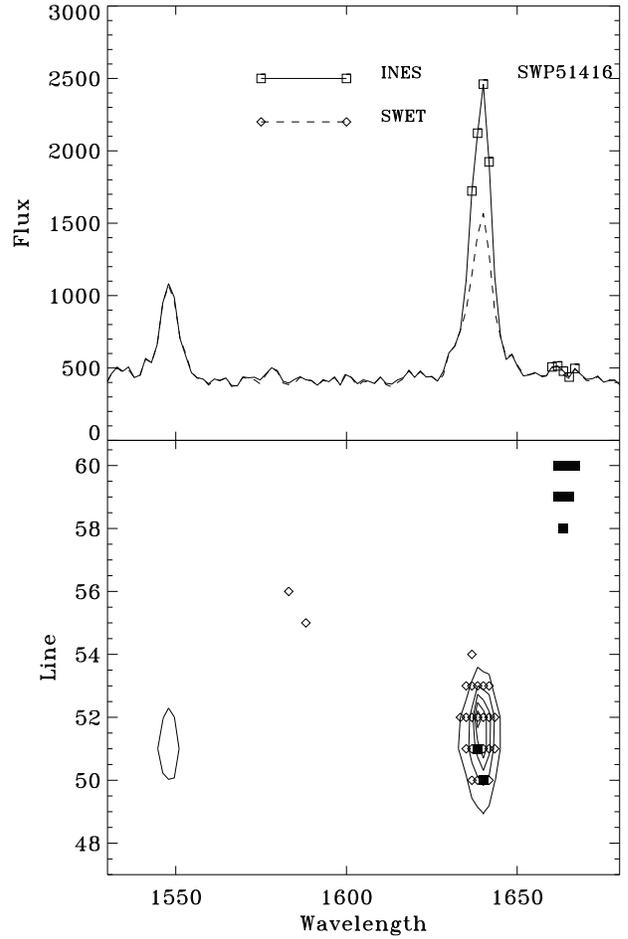,width=9.cm}
\caption{This figure is similar to Figure \ref{fig:fl1}, but in 
this spectrum the HeII line is flagged in the \ines\ spectrum, while it is
not in the \swet\ extraction.  As shown in the bottom panel, there are two
pixels flagged as saturated in the SILO file.  While these flags are
propagated to the \ines\ extracted spectrum, they disappear in
the \swet\ extraction. Symbols are as in Figure \ref{fig:fl1}. In
particular, the two solid squares on the HeII line in the SILO file mark
saturated pixels.}
\label{fig:fl2}
\end{center}
\end{figure}
  
One of the advantages of optimal extraction techniques is that they
should be able to recover the flux at flagged pixels as far as the correct
cross-dispersion profile is used. However, this ability must be analyzed
carefully since flagged pixels are already excluded from the determination
of the weighting profile. As an example we will discuss the case of an
emission lines spectrum.

In many cases the exposure times were chosen to get a good exposure level
in the continuum, frequently resulting in saturation for the the peak of
the lines.  Then, the core of the strongest lines are flagged as
"Extrapolated ITF" or "saturated". If there are only a few pixels flagged
it is expected that the correct flux will be obtained from a correct
profile. However, the beam pulling effect in \iue\ images shifts the strong
lines with respect to the continuum. Even in the case that the method would
be able to reproduce such short scale shifts, if flagged pixels are not
used to determine the spatial profile, the weighting profile will be
shifted with respect to the actual spatial profile of the line that will be
treated as a cosmic ray. For this reason, in the \ines\ extraction only
pixels with no real flux information are discarded: reseaux marks, pixels
not photometrically corrected, 159DN corrupted pixels and telemetry
dropouts.

The way the information about bad quality pixels is passed onto the final
1-D output spectrum is also related to the role these pixels play in the
extraction procedure. In the \ines\ extraction, a conservative approach has
been followed and the flag of any pixel in the SILO file that makes a
contribution to the final 1-D extracted spectrum (i.e. for which the
extraction profile is not zero) is passed into the 1-D flag spectrum. This
method may propagate flags of pixels whose contribution is almost
negligible (e.g. reseaux marks within the aperture, but outside the PSF),
but assures that no relevant quality flag is lost. Figure~\ref{fig:fl2}
illustrates a case where there are two pixels in the HeII line with the
saturation flag in the SILO file. \swet\ treats the line pixels as a
"cosmic rays" (note the "-32" flags in SILO file), but neither these flags
nor the saturation flags are passed onto the final 1-D spectrum. In
contrast, \ines\ reproduces the correct flux and flags the wavelength bins
where there are saturated pixels.

    	\subsubsection{Solar contamination removal}
	\label{sec:solarext}

\begin{figure}
\begin{center}
\epsfig{file=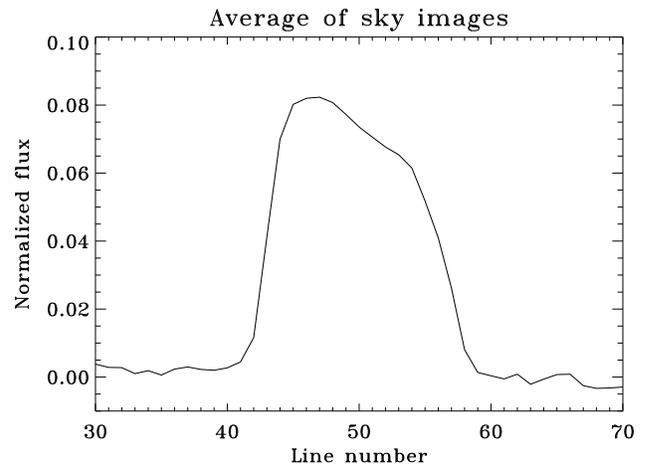,width=9.cm}
\caption{Average spatial profile of solar scattered light, based on sky 
exposures}
\label{fig:sunprof}
\end{center}
\end{figure}

By the end of its operational life, the \iue\ telescope was affected by the
so-called FES anomaly (P\'erez and Pepoy 1997). In reality, it was not an
anomaly of the FES functionality but that name was given because the
problem was firstly detected on FES images (Rodr\'{\i}guez-Pascual
1993). For an unknown reason, scattered Sun and Earth light was entering
the telescope tube and reaching the on-board detectors (FES and SEC Vidicon
cameras).  On FES images this light was known as the ``streak'' because it
filled only a portion of the image, producing a pseudo-background. Under
the worst conditions the FES detector was fully saturated, providing a number of
counts similar to that from a 5th magnitude star.  The analysis of the
problem showed that light scattered into the telescope was mainly solar in origin
(Rodr\'{\i}guez \&\ Fernley 1993). The effect on science images was to
contaminate LWP low resolution images with an extended spectrum filling the
whole aperture (Fig.~\ref{fig:sunprof}). SWP images were not affected
because of the solar-like spectrum of the scattered light and no measurable
contamination has been detected in LWP high resolution spectra.  Two types
of contamination were identified in LWP images, depending on whether the dominant
source was direct sunlight or sunlight reflected on the Earth
(Rodr\'{\i}guez-Pascual
\&\ Fernley 1993).

\begin{figure}
\begin{center}
\epsfig{file=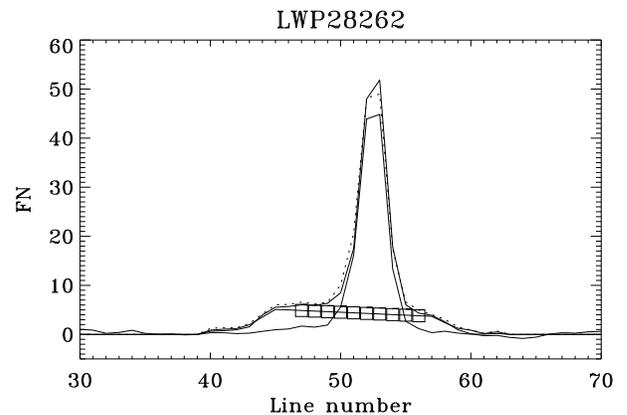,width=9.cm}
\caption{The thick line shows the average spatial profile 
from 2900\AA\ to 3300\AA\ in the image LWP28262. The contribution from the
solar scattered light contamination and the extraction
profile used for the point source are shown as thin lines and the sum of
both is represented by crosses. 
Squares show the extrapolated points of the solar contribution.
Note that there is still some remnant
contribution of the extended source in the left side of the point source
profile.}
\label{fig:profex}
\end{center}
\end{figure}

LWP images contaminated with solar scattered light are identified 
as extended sources by \newsips . However, the \swet\ extraction module is
forced to perform a point-like source extraction, i.e., restricted to 13
spectral lines, in all LWP images taken after November 1992 and whose IUECLASS 
does not correspond to solar system objects or sky exposures. This
approach does not reduce the solar contribution to the extracted spectrum
in a consistent way and definitely does not remove it completely.

Several methods have been evaluated to correct this contamination.  The
correlation between the strength of the streak as measured with the FES and
the strength of the contamination in spectral images led to consider the
possibility of building up a spectral template to be scaled by the FES
counts. However, this approach was not useful in practice because of the
two types of solar spectra found and the large scatter in the FES
counts-spectral flux relation (Rodr\'{\i}guez-Pascual \&\ Fernley 1993)
associated with the specific light scattering geometry.

The procedure developed in the \ines\ extraction was designed to handle
only the most straightforward case: a point-like source, well centered into
the aperture. The spectrum of the target does not fill the whole aperture
and the solar contamination can be estimated from the spectral lines on
both sides of the target PSF. Obviously, this method only works on large
aperture spectra; contamination in the small aperture is not corrected.

\begin{figure}
\begin{center}
\epsfig{file=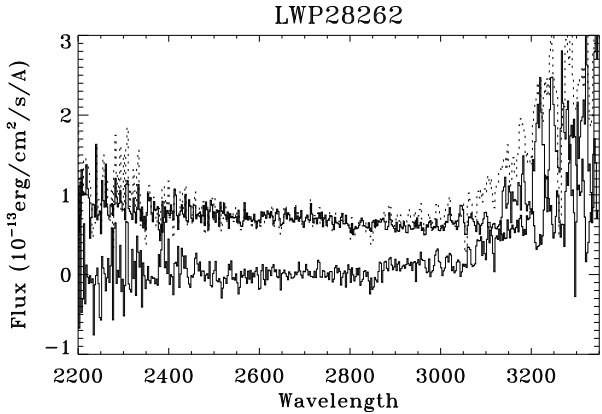,width=9.cm}
\caption{This figure illustrates
how the \ines\ extraction software is able to get rid of most of the solar
scattered light contamination. The thick line is the spectrum of the point
source, the thin line represents the solar spectrum and the dotted line is
a direct boxcar extraction of the whole aperture.}
\label{fig:solarex}
\end{center}
\end{figure}

The first step is to identify whether an image is affected by solar
contamination. The check is done only on LWP images taken after November
1992 since this was the time its presence was first detected in the
FES. The procedure searches for the peak of the average spatial profile.
If the contribution of a point source, i.e. up to 11 spectral lines wide,
is between 5\% and 95\% of the total spatial profile then the presence of
both an extended and a point-like sources is assumed. This method obviously
does not guarantee that the extended source is due to the solar
contamination; it may happen that the observation corresponds to a crowded
field with several sources. However, we have adopted this approach because
any potential user of crowded fields data should already be aware that the
\iue\ Project does not provide individual spectra when several sources are
within the aperture. Such spectra need to be individually analyzed from the
SILO file. But any user interested in the archival data of an isolated
object should not have to worry about contamination by other sources and
can take the extracted spectra as the real spectra of the isolated
object. Bearing this in mind, it was decided to accept the risk that in
some cases the procedure will remove the contribution of an extended
component that is not the solar contamination.

Once a LWP image has been identified as contaminated, the 2-D spectrum of
the solar light is reconstructed.  First, the solar spectrum is extracted
as in the standard case, but masking out 11 spectral lines centered at the
location of the peak in the average spatial profile.  Since sky exposures
show that the cross-dispersion profile of the solar contamination is
roughly linear in the center of the aperture (Fig.~\ref{fig:sunprof}), the
2-D spatial profile of the solar contamination within the point-like source
location is derived interpolating linearly from the wings of the
profile. The 2-D contamination is then reconstructed and subtracted from
the SILO file.  The point-like spectrum is extracted from the resulting
corrected SILO following the standard \ines\ procedures.  Spectra in which
the correction for solar contamination has been applied are identified by
the following message in the FITS header: {\tt *** WARNING: SOLAR
CONTAMINATION CORRECTION APPLIED}

In figures \ref{fig:profex} and \ref{fig:solarex} we show an example of the
performance of the method. The average spatial
profile in the range 2900-3300\AA\ is shown as a thick line in
Fig~\ref{fig:profex}; the thin lines show the profiles estimated for the
extended and point sources (crosses represent the sum of both). The squares
show the spectral lines discarded to estimate the extended source and later
interpolated. The corresponding output spectra are shown in 
Fig~\ref{fig:solarex}.
 
The performance of this technique has been tested using the data of the
blazar PKS~2155-304, extensively monitored with \iue . In particular, two
intensive monitoring campaign were carried out in 1991 and 1994 (Pian et
al., 1997), before and after the appearance of the FES anomaly.  During the
1991 campaign, 98 LWP spectra were obtained. In 1994, \iue\ was
continuously pointing to this target for 10 days starting on May 15th. A
total of 236 spectra were obtained, half of them with the LWP
camera. Albeit the flux level of the target varied between both runs and
even within each run, the effect of the solar scattered light into the LWP
camera can be tested because the changes in the spectral shape are small
(Pian et al. 1997).

\begin{figure}
\psfig{file=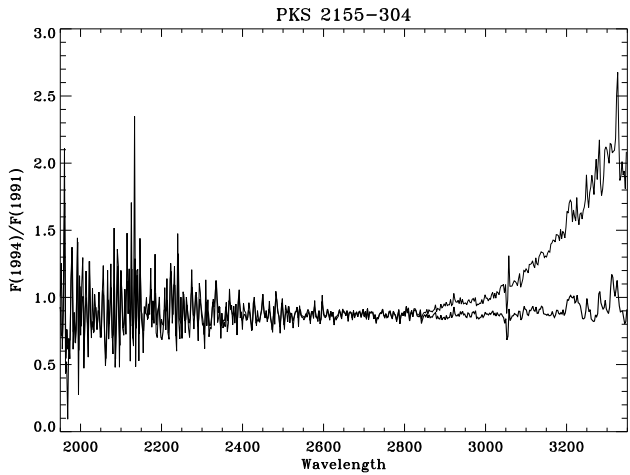,width=9.cm}
\caption{Ratio between the average 
spectra of PKS~2155-304 during the observing campaigns in 1991 and
1994. The thin line is the ratio between \swet\ spectra and the thick line
shows the ratio between INES spectra. Although the 1994 flux is on average
10\% lower than in 1991, the index of the power law describing the UV
spectrum has not changed noticeably, as indicated by the constant ratio
below 2800\AA. The sharp rise of the \swet\ ratio is due to the solar
contamination.}
\label{fig:pksave}
\end{figure}

First we compare the ratio of the \swet\ average spectra of both campaigns
(Fig~\ref{fig:pksave}). This ratio shows a sharp turn-up beyond 2800\AA\
due to the solar contamination, but the ratio of the \ines\ averages is
essentially independent of the wavelength.  This is a clear demonstration
that \swet\ is not able to remove the scattered light in the output
spectrum. The features beyond 3200\AA\ are typically  due to the low S/N
in this region of the \iue\ LWP camera in the individual spectra.

\begin{figure}
\psfig{file=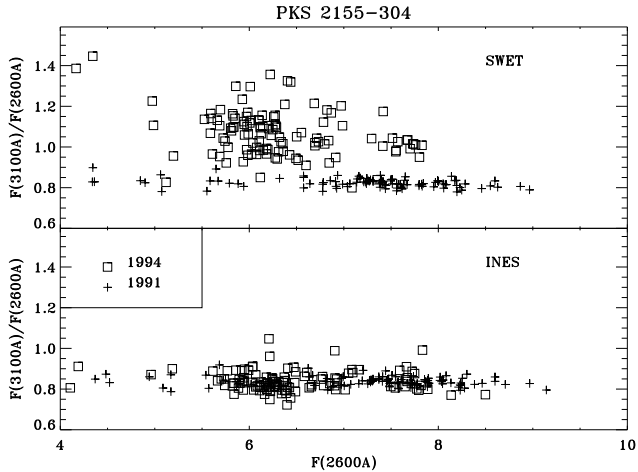,width=9.cm}
\caption{The 3100\AA\ region in the \swet\ extracted spectra of 
PKS~2155-304 ({\it upper panel}) is strongly contaminated by solar scattered
light as shown by the ratio F(3100\AA)/F(2600\AA) in 1994 ({\it squares})
and 1991 ({\it crosses}). The values of this ratio during 1994 and 1991 are
in much better agreement when \ines\ extracted spectra are used ({\it lower 
panel})}
\label{fig:pksrat}
\end{figure}

Another test of the presence of solar scattered light in the output spectra
is to compare the ratios of fluxes in different wavelength bands.  For each
campaign and extraction method we have compared the relation between the
flux at 2600\AA, where no solar contamination is expected and the ratio of
the fluxes at 3100\AA\ and 2600\AA. This ratio can be taken as a measure of
the amount of contamination since the band centered at 3100\AA\ is the most
affected. The results are shown in Fig~\ref{fig:pksrat}. The
F(3100\AA)/F(2600\AA) ratio is definitely larger for the 1994 spectra
extracted with \swet . However, the 1991 and 1994 values of this ratio for 
\ines\ spectra are indistinguisable, although there are still a
few data points for which the ratio is larger by $\sim$20\%.

\subsection{Homogenization of the wavelength scale}
	\label{sec:homwave}

One of the main purposes of the modifications implemented within the \ines\
system is to provide the data in such a form that the user needs to perform
the minimum number of operations before starting the scientific analysis
and to decrease the instrumental dependence of the extracted spectrum
(important for further use by scientist without specific \iue\ knowledge).

One of the characteristics of the \swet\ low resolution spectra which,
although well documented, can originate some confusion to users, is that
the low resolution long wavelength data do not have an uniform wavelength
scale, i.e., there are long wavelength spectra with different stepsize and
with different number of points in the extracted data.  These differences
depend on the date of observation and, in the case of the LWR camera, on
the ITF used in the processing. The dependency on the date of observation
is very small (and it is also present in the SWP camera), but differences
between both long wavelength cameras and between both LWR ITFs cannot be
neglected. The difference is only in the size of the wavelength step and
not in the starting wavelength of the NET spectra (1050 and 1750 \AA\ for
short and long wavelength ranges, respectively). Since the Inverse
Sensitivity curves are not defined for the full spectral range of the
extracted data, LWP and LWR \newsips\ low resolution {\it calibrated} spectra
do not start at the same wavelength and have a different number of points.

Any combination or comparison of long wavelength spectra would require the
rebinning to a common wavelength scale. In order to facilitate the use of
the extracted data, this rebinning has already been built in the \ines\ processing
system, assuring homogeneity in the data.

Table \ref{tab:step} summarizes the low resolution wavelength step used in
\newsips\ for each camera.
 
\begin{table}[htb]
\begin{center}
\caption[NEWSIPS Wavelength Steps]{\em Summary of the wavelength steps of 
\newsips\ low resolution spectra}
\vspace*{0.3cm}
\footnotesize
\begin{tabular}{l c c c }
\hline
Camera & $\lambda$ step & First calibrated & \#\ of calibrated \\ & (\AA) &
pixel & pixels \\
\hline
LWP			&		&			&		\\
pre-1990	      	& 2.6627	&	1851.181	&	563	\\	
post-1990	      	& 2.6628    	&	1851.186	&	563	\\
			&		&			&		\\
LWR-ITF A               & 		&			&		\\
pre-1980.1  		& 2.6658	&	1851.300	&	563	\\
post-1980.1 		& 2.6657	&	1851.298	&	563	\\
			&		&			&		\\
LWR-ITF B               &		&			&		\\
pre- 1979.9  		& 2.6693	&	1851.433	&	562	\\
post- 1979.9 		& 2.6689	&	1851.418	&	562	\\
			&		&			&		\\
SWP                     & 		&			&		\\	
pre-1990	      	& 1.6763	&	1150.578	&	495	\\
post-1990	      	& 1.6764	&	1150.584	&	495	\\
\hline

\end{tabular}
\normalsize
\label{tab:step}
\end{center}
\end{table}

The resampling was performed following this approach:

\begin{itemize}

\item	 All the long wavelength spectra are rebinned to the same
wavelength step.

\item	 The size of the new wavelength step is taken as  the largest one of all 
the steps used, i.e. 2.6693 \AA\ per pixel for the LW cameras, and 1.6764
\AA\ per pixel for SWP. The number of calibrated pixels is 495 for the SWP
camera, and 562 for the long wavelengths cameras.

\item 	 The starting wavelength of the calibrated spectrum has not been 
modified (i.e. it is the first pixel within the spectral region in which
the Inverse Sensitivity Curves are defined). 

\item	Only flux-calibrated points are included in the final spectrum.

\item	Both the absolute flux and the sigma spectra are rebinned.   
 
\item   The rebinning of the flux spectrum is made through the following 
expression: 
 
\begin{equation} 
F_i = \frac {\sum_j w_j f_j} {\sum_j w_j} 
\end{equation} 
 
where F$_i$ is the flux of the final rebinned pixel, f$_j$ are the fluxes
of the input pixels, and $\omega_j$ are the fractions of each original pixel
within the new one. It must be noted that for this particular case in which
the original and the final wavelength steps are very close to each other,
this procedure provides results very similar to a simple linear
interpolation.

\item	 The procedure to handle the sigma spectrum is similar, but using
the square of the errors instead:	
 
\begin{equation} 
E_i = \sqrt{\frac {\sum_j w_j e^2_j} {\sum_j w_j}}
\end{equation} 

In this expression, E$_i$ and e$_j$ are the rebinned and original errors,
respectively.

\item	 The $\nu$-spectrum is also  re-computed. Each final pixel
has the minimum (i.e. the ``worst'') $\nu$ of the original pixels used in
the rebinning.  The number of flags in the rebinned spectrum is larger than
in the original one, since every pixel contributes to at least two final
pixels (e.g. a reseau mark originally flagged in two consecutive pixels
would result in three flagged points in the resampled spectrum).

It must be noted that the largest increase in the size of wavelength step
(which corresponds to the LWP camera) is by a 0.25\%. The spectral
resolution for this camera is 5.2\AA (1.95 pixels, Garhart et
al. 1997). Consequently this rebinning does not introduce any significant
degradation in spectral resolution.

\end{itemize}

\section{{\em INES} Data quality evaluation}
	\label{sec:dataev}

	\subsection{Flux repeatability}

The repeatability of the \ines\ low resolution spectra has been tested on a
large sample of spectra of some of the IUE standard stars.  The only
restriction imposed has been to include only non--saturated spectra of
similar level of exposure (i.e. similar exposure times) in order to avoid
the remaining non--linearity effects (see Section \ref{sec:lin}). The
spectra cover all the range of observing epochs and camera
temperatures. Therefore it must be taken into account that the
repeatability, as defined here, includes implicitly the uncertainties in
the camera time degradation and the temperature corrections.

The study has been performed in 100 \AA\ wide bands. Table~\ref{tab:rep} lists 
the central wavelength of the bands and the repeatability,
defined as the percent rms respect to the mean intensity of the band. The
figures in brackets are the number of spectra considered in each case.

\begin{table}
\caption{Flux repeatability of INES low resolution spectra}
\centerline{SWP}
\scriptsize
\begin{center}
\begin{tabular}{l c c c c}
\hline
	& BD+28~4211 & BD+75~325 & HD~60753 &   Average \\
 Band	& (292)      &  (196)    &  (228)   &		\\
\hline 
  1200  &  3.82   &   3.11   &    5.29  &  4.07 \\
  1300  &  3.12	  &   2.50   &    3.08  &  2.90 \\
  1400  &  1.99	  &   1.95   &    2.09  &  2.01 \\
  1500  &  1.84	  &   1.70   &    1.90  &  1.81 \\
  1600  &  2.00	  &   1.86   &    2.10  &  2.04 \\
  1700  &  2.21	  &   1.68   &    1.99  &  1.96 \\
  1800  &  2.03	  &   1.74   &    1.97  &  1.91 \\
  1900  &  2.10	  &   2.03   &    1.94  &  2.02 \\
\hline
\end{tabular}
\end{center}
\normalsize

\centerline{LWP}
\scriptsize
\begin{center}
\begin{tabular}{l c c c c}
\hline
      & BD+28~4211 & BD+75~325 & HD~60753 &  Average \\
Band  &  (232)     &   (223)   &    (225) &          \\
\hline
  1900  &  18.21    &   16.90   &    15.69   &  16.93 \\
  2000  &   3.43    &    3.65   &     4.75   &   3.94 \\
  2100  &   4.19    &    3.81   &     4.61   &   4.20 \\
  2200  &   3.11    &    3.29   &     4.83   &   3.74 \\
  2300  &   3.54    &    3.59   &     4.49   &   3.87 \\
  2400  &   2.78    &    2.71   &     3.19   &   2.89 \\
  2500  &   2.35    &    2.38   &     2.63   &   2.45 \\
  2600  &   2.21    &    2.17   &     2.64   &   2.34 \\
  2700  &   1.99    &    2.13   &     2.27   &   2.23 \\
  2800  &   1.94    &    2.07   &     2.06   &   2.02 \\
  2900  &   1.94    &    2.26   &     2.06   &   2.09 \\
  3000  &   2.42    &    2.71   &     2.24   &   2.46 \\
  3100  &   4.00    &    4.12   &     3.51   &   3.88 \\
  3200  &   7.77    &    6.93   &     5.64   &   6.78 \\
  3300  &  14.90    &   16.84   &    11.29   &  14.34 \\
\hline
\end{tabular}
\end{center}
\normalsize

\centerline{LWR}
\scriptsize
\begin{center}
\begin{tabular}{l c c c c}
\hline
        & BD+28~4211 & BD+75~325 & HD~60753   & Average \\
 Band   &  (89)      &  (86)     &   (76)     &         \\
\hline
  1900  &   7.77   &    7.14    &      7.92   &   7.61  \\
  2000  &   4.51   &    4.99    &      5.66   &   5.05  \\
  2100  &   3.84   &    4.25    &      5.26   &   4.45  \\
  2200  &   4.07   &    4.72    &      5.50   &   4.76 \\ 
  2300  &   3.71   &    3.64    &      4.57   &   3.97 \\ 
  2400  &   3.43   &    3.48    &      4.24   &   3.72 \\
  2500  &   2.92   &    3.00    &      3.19   &   3.04 \\
  2600  &   3.05   &    2.94    &      3.55   &   3.18 \\
  2700  &   3.20   &    3.18    &      3.95   &   3.44 \\
  2800  &   3.42   &    3.66    &      3.66   &   3.58 \\
  2900  &   3.67   &    3.77    &      3.77   &   3.74 \\
  3000  &   4.31   &    3.19    &      4.17   &   3.89 \\
  3100  &   5.18   &    4.28    &      4.62   &   5.69 \\
  3200  &  10.19   &    9.18    &     10.22   &   9.86 \\
  3300  &  50.35   &   41.19    &     54.17   &  48.57 \\
\hline
\end{tabular}
\end{center}
\normalsize
\label{tab:rep}
\end{table}

As expected, the best repeatability is attained in the regions of maximum
sensitivity of the cameras. In the SWP the repeatability is around 2\%
longward 1400 \AA. For the LWP camera, values lower than 3\% are reached in
the central part of the camera, 2400-3000 \AA.  At the extreme wavelengths
the repeatability is around 15\%. The results are slightly worse for LWR
most likely due to the instability of the camera after it ceased to be used
for routine operations. The repeatability is between 3-4\% in the region
2300-3000 \AA. Particularly bad is the 3300 \AA\ band, but at
the shortest wavelengths (1850-1950 \AA) the repeatability is substantially
better than in the LWP camera. When considering only images taken when LWR
was the prime long wavelength camera, the repeatability is similar to that
of LWP in the central part of the camera.

  	\subsection{Reliability of extraction errors}

\begin{figure}[hbt]
\psfig{file=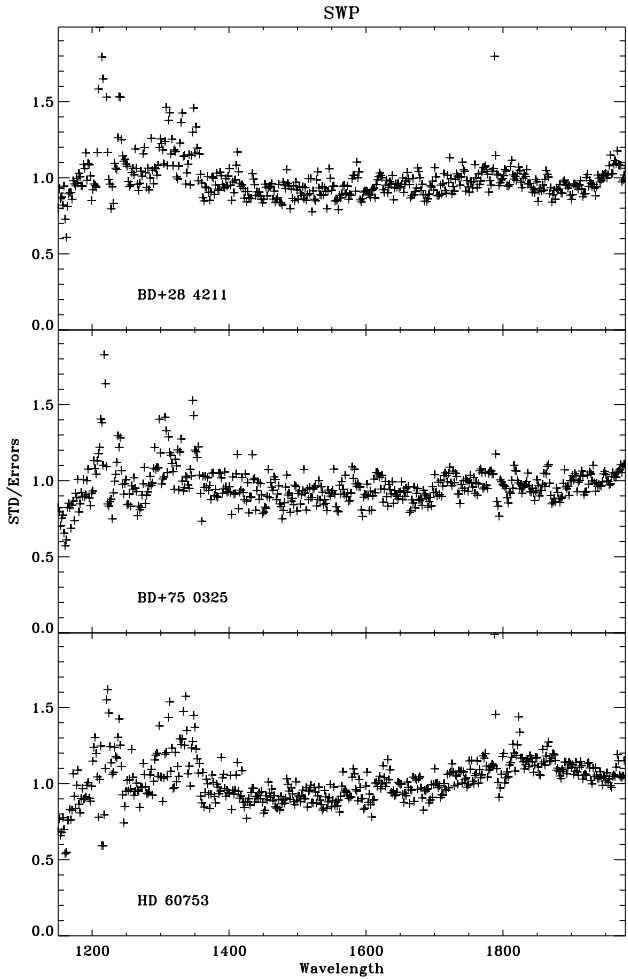,width=9cm} 
\caption{Comparison of the extraction errors ({\em Errors}) in INES 
with the dispersion around the mean spectrum ({\em STD}) for three standard
stars, after correction with the coefficients shown in Table~\ref{tab:err}
for the SWP camera.}
\label{fig:errswp} 
\end{figure}

\begin{figure}[hbt]
\psfig{file=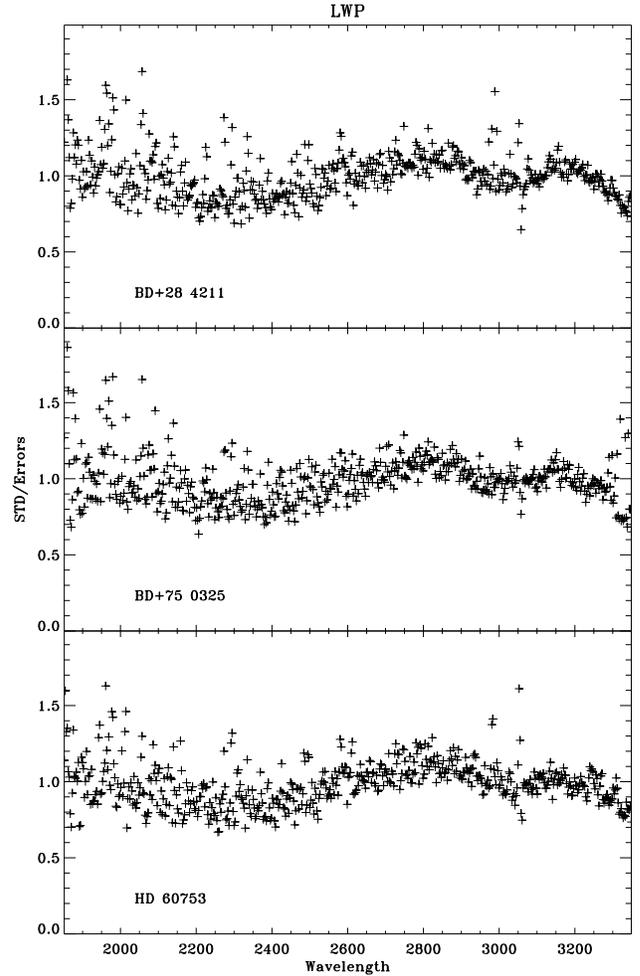,width=9cm} 
\caption{Same as Figure \ref{fig:errswp}, but for the LWP camera}
\label{fig:errlwp} 
\end{figure} 

In addition to the flux spectrum, optimal extraction methods also provide
an error spectrum. Formally, these errors only account for the
uncertainties in the extraction procedure, based on the noise model of the
detector. 
They do not include
uncertainties  driven by parameters affecting the image registration. 
During the processing, corrections are applied to account for
the changes in temperature in the head amplifier of the cameras 
(THDA) and the loss of sensitivity of the detectors. All these corrections 
have their own uncertainties. There are yet other systematic errors
that affect the absolute fluxes, as the uncertainty in the inverse
sensitivity curve, but do not affect the comparison of different sets of
\iue\ spectra. 
The extraction errors can be used to compare fluxes in different bands of
the same spectrum or to compute weighted averages of a set of spectra, but
they may not be appropriate to evaluate the variability of a source or an
spectral feature, due to the considerations given above.

\begin{figure}[hbt]
\psfig{file=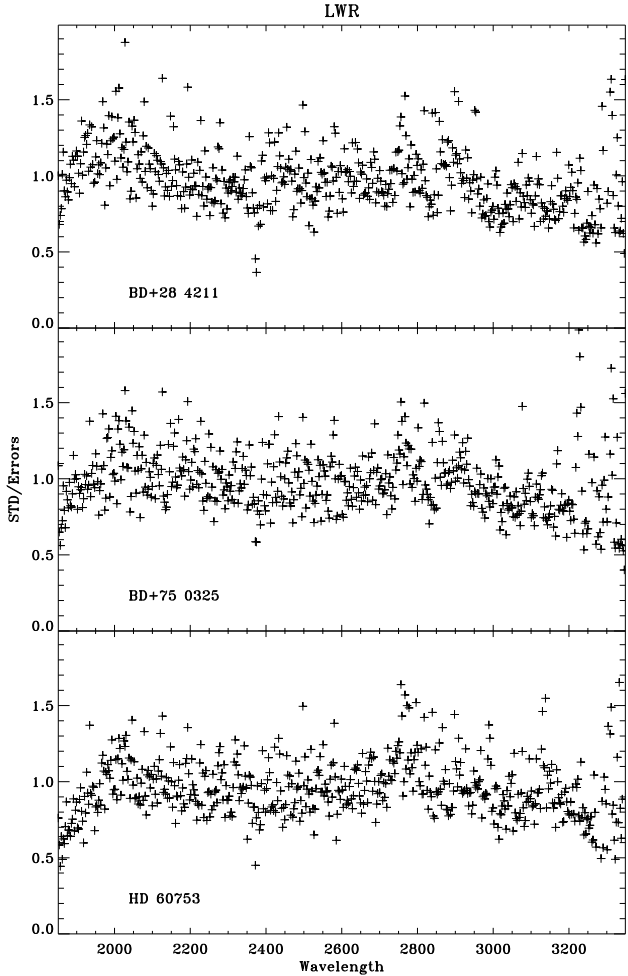,width=9cm} 
\caption{Same as Figure \ref{fig:errswp}, but for the LWR camera with a
voltage of -5.0 kV.}
\label{fig:errlwr} 
\end{figure}

In order to check the statistical validity of the errors provided by the
\ines\ extraction, we have taken the same data set used in the previous
section (i.e. a large sample of spectra of standard stars with similar
level of exposure) and compared the rms around the mean with the average
errors as given by the extraction procedure.  In general, the extraction
errors underestimate the errors represented by the rms in the three
cameras, with the exception of the shortest wavelength end of the LWP
camera. In the SWP camera the errors are underestimated by $\sim$20-40\% ,
depending on the wavelength. In the LWR the ratio between extraction and
actual errors is nearly constant (12\%) all along the camera, while in the
LWP the discrepancy can be as large as 40\%\ at the longest wavelength. In
this camera, shortward 2400\AA\ the extractions errors are too large by
15--20\%. This region is very noisy and there are reasons to suspect that
such a noise departs significantly from a gaussian behaviour.

It is also found that the dependency of the ratio STD/Error (where ``STD''
is the standard deviation around the mean spectrum, and ``Error'' is
derived from the extraction errors) with wavelength can be well
represented by a straight line with the coefficients shown in
Table~\ref{tab:err}. Reliable values for these
coefficients could not be obtained for the LWR camera operated at -4.5kV
due to the scarcity of data.
\begin{table}
\caption{}
\begin{center}
\begin{tabular}{lcc} 
\hline

Camera & a & b  \\ 
\hline
 LWP         & -0.49 & 6.2$\times10^{-4}$ \\ 
 SWP         & 1.11  & 1.6$\times10^{-4}$ \\ 
 LWR(-5.0kV) & 1.12  & -3.1$\times10^{-6}$ \\ 
\hline
\end{tabular}
\end{center}
\label{tab:err}
\end{table}

In order to compare fluxes in different spectra of the same object, the
extraction errors must be modified according to
\begin{equation}
 \varepsilon(\lambda)=a+b\varepsilon_{{\rm E}}(\lambda)
\end{equation}
where $a$\ and $b$\ are the coefficients in Table~\ref{tab:err} and 
$\varepsilon_{{\rm E}}(\lambda)$\ are the extraction errors.

The results of the application of this correction are shown in figures 
\ref{fig:errswp}, \ref{fig:errlwp} and \ref{fig:errlwr}. The dispersion
around the expected value of 1 is 0.15 for SWP, 0.17 for LWP and 0.18 for
LWR. The structure still seen in these figures might be related to
remaining non linearities in the ITF's (see below).

  	\subsection{Flux Linearity}
	\label{sec:lin}

Despite the correction applied during the processing of the \iue\ data
through the application of the Intensity Transfer Functions (ITFs), the
final spectra are still affected by non-linearities to some degree. As a
consequence, spectra of the same non-variable object observed with
different exposure times might have slightly different flux levels.

\begin{figure*}
\epsfig{file=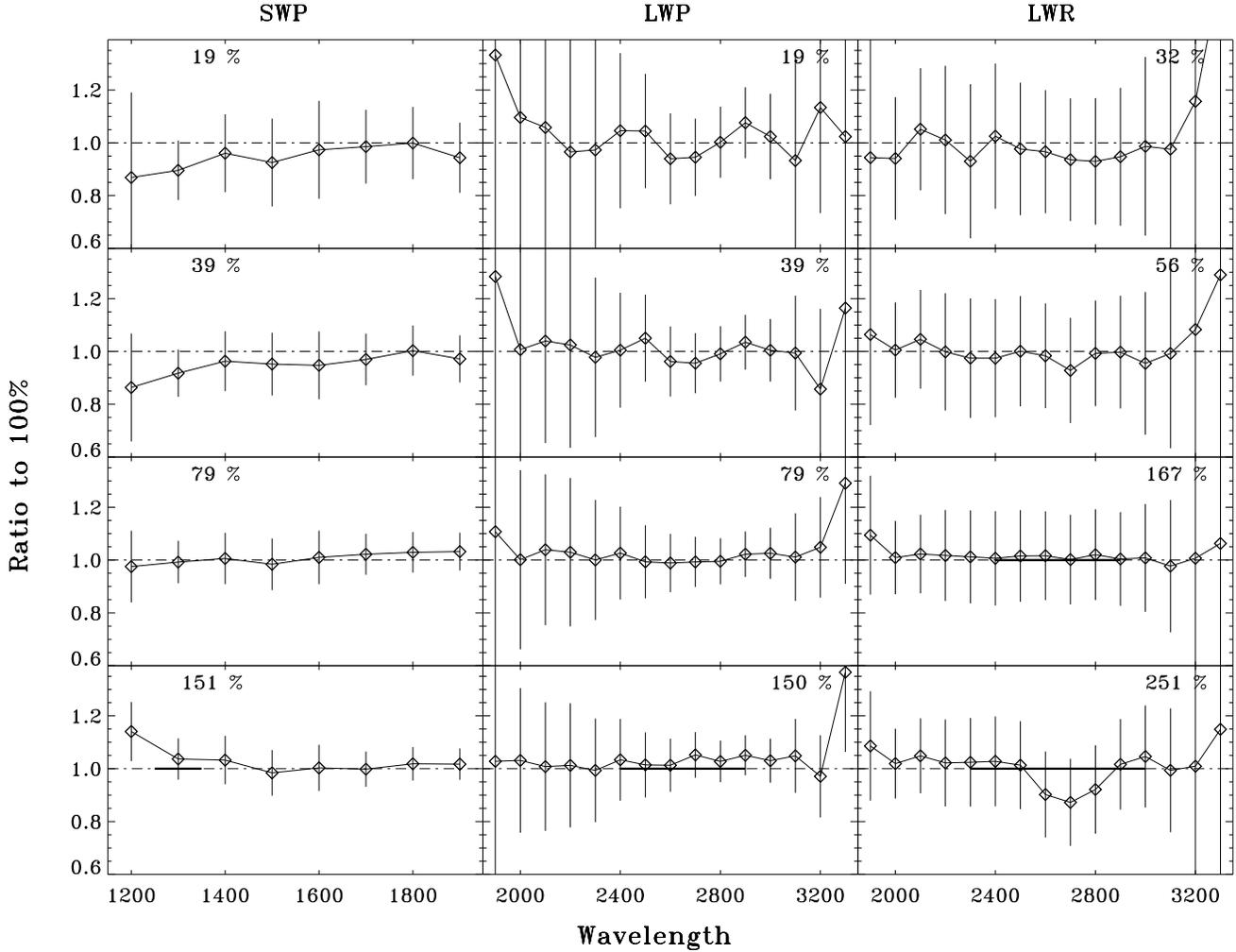,height=18cm,angle=90}
\caption{Ratios of different exposure levels to the
100\%\ for the three cameras. The thick horizontal line marks the saturated part 
in the overexposed spectra}
\label{fig:alllin}
\end{figure*}


\begin{table}
\caption{Linearity properties of INES low resolution spectra (ratios to
100\%\ exposure)}

\centerline{SWP}
\begin{center}
\scriptsize
\begin{tabular}[h]{l c c c c c c c c}
\hline
Band & 8\% & 19\% & 28\% & 39\% & 59\% & 79\% & 121\% & 151\% \\
\hline
1200 & 0.78 & 0.87 & 0.88 & 0.86 & 0.92 & 0.98 & 1.04 & 1.14\\
1300 & 0.81 & 0.90 & 0.87 & 0.92 & 0.96 & 0.99 & 1.03 & 1.04 \\
1400 & 0.92 & 0.96 & 0.95 & 0.96 & 0.96 & 1.01 & 1.02 & 1.03 \\
1500 & 0.88 & 0.93 & 0.93 & 0.95 & 0.96 & 0.98 & 0.99 & 0.98 \\
1600 & 0.92 & 0.97 & 0.96 & 0.95 & 0.98 & 1.01 & 1.02 & 1.00 \\
1700 & 0.83 & 0.99 & 0.97 & 0.97 & 1.00 & 1.02 & 1.03 & 1.00 \\
1800 & 0.98 & 1.00 & 0.97 & 1.00 & 1.00 & 1.03 & 1.01 & 1.02 \\
1900 & 0.96 & 0.94 & 0.95 & 0.97 & 1.00 & 1.03 & 1.03 & 1.02 \\
\hline
\end{tabular}
\normalsize
\end{center}

\centerline{LWP}
\begin{center}
\scriptsize
\begin{tabular}[h]{l c c c c c c c}
\hline
Band &      19\% 	 	 &    39\% 	   &    59\% 	     &
79\% 	       &   129\% 	&   150\% 	   &   200\% \\
\hline
1900 & 1.33 & 1.28 & 1.04 & 1.11 & 1.03 & 1.03 & 1.01 \\
2000 & 1.10 & 1.01 & 1.01 & 1.00 & 1.02 & 1.03 & 1.00 \\
2100 & 1.06 & 1.04 & 1.03 & 1.04 & 1.02 & 1.01 & 1.04 \\
2200 & 0.97 & 1.02 & 0.99 & 1.03 & 1.02 & 1.01 & 1.06 \\
2300 & 0.97 & 0.98 & 1.00 & 1.00 & 1.02 & 0.99 & 1.01   \\
2400 & 1.05 & 1.00 & 1.03 & 1.03 & 1.04 & 1.03 & 1.05   \\
2500 & 1.04 & 1.05 & 0.99 & 0.99 & 1.00 & 1.01 & 1.07 \\
2600 & 0.94 & 0.96 & 0.96 & 0.99 & 1.00 & 1.01 & 1.09 \\
2700 & 0.95 & 0.96 & 0.99 & 0.99 & 1.02 & 1.05 & 1.12 \\
2800 & 1.00 & 0.99 & 1.02 & 1.00 & 1.03 & 1.03 & 1.11 \\
2900 & 1.08 & 1.04 & 1.05 & 1.02 & 1.05 & 1.05 & 1.08 \\
3000 & 1.02 & 1.00 & 1.02 & 1.03 & 1.05 & 1.03 & 1.05  \\
3100 & 0.93 & 0.99 & 0.95 & 1.01 & 1.07 & 1.05 & 1.02 \\
3200 & 1.13 & 0.86 & 0.95 & 1.05 & 1.10 & 0.97 & 1.07 \\
3300 & 1.02 & 1.16 & 1.30 & 1.29 & 1.11 & 1.37 & 1.12 \\
\hline
\end{tabular}
\normalsize
\end{center}

\centerline{LWR-ITF-B}
\begin{center}
\scriptsize
\begin{tabular}[h]{l c c c c}
\hline
Band  &     32\% &    56\%  	  &  167\%	&  251\% \\
\hline 
 1900 & 0.94 & 1.06 & 1.09 & 1.09 \\
 2000 & 0.94 & 1.01 & 1.01 & 1.02 \\
 2100 & 1.05 & 1.05 & 1.02 & 1.05 \\
 2200 & 1.01 & 1.00 & 1.02 & 1.02 \\
 2300 & 0.93 & 0.97 & 1.01 & 1.02 \\
 2400 & 1.03 & 0.97 & 1.01 & 1.03 \\
 2500 & 0.98 & 1.00 & 1.02 & 1.01 \\
 2600 & 0.97 & 0.98 & 1.02 & 0.90 \\
 2700 & 0.94 & 0.93 & 1.00 & 0.87 \\
 2800 & 0.93 & 0.99 & 1.02 & 0.92 \\
 2900 & 0.95 & 1.00 & 1.00 & 1.02 \\
 3000 & 0.99 & 0.95 & 1.01 & 1.05 \\
 3100 & 0.98 & 0.99 & 0.98 & 0.99 \\
 3200 & 1.16 & 1.08 & 1.01 & 1.01 \\
 3300 & 1.63 & 1.29 & 1.06 & 1.15 \\
\hline
\end{tabular}
\end{center}
\label{tab:lin}
\end{table}

In order to evaluate the importance of the remaining non-linearities we
have chosen, for each camera, a set of low resolution spectra of the
standard star BD+28~4211, extracted with the \ines\ system, obtained
close in time under similar temperature conditions and with different
exposure times. In each set one of the spectra is defined as a 100\%\
exposure, and all the other are referred to that one.

The summary of the data  used for each camera is as follows:

\begin{itemize}

\item SWP: Nine spectra taken in December 1993, with exposure times
ranging from 2 sec (9\%) to 40 sec (150\%).

\item LWP: Nine spectra taken in October 1986, with exposure times
ranging from 10 sec (20\%) to 100 sec (200\%).

\item LWR: 	Five spectra taken in August 1980, with exposure times
ranging from 20 sec (30\%) to 150 sec (250\%). All this spectra have been
processed with ITF--B (Garhart et al. 1997), which is the one giving the best correlation
coefficient in this case, as in most of the pre-1984 LWR spectra.

\end{itemize}

Each spectrum was binned into in 100 \AA\ bands and divided by the
corresponding reference exposure.  The results are summarized in Table
\ref{tab:lin}.  Examples of the behaviour of different spectral bands for
each of the cameras as a function of the level of exposure are shown in
Figure
\ref{fig:alllin}.

In the SWP camera the largest departures from linearity are found at the
short wavelength end of the underexposed spectra, where flux can be
underestimated by up to a 20\%.  Apart from this case, longward Lyman
$\alpha$ ratios to the 100\% spectrum are generally within$\pm$5\%. The
best results are achieved in the 1800 \AA\ band, where linearity is within
$\pm$ 3\%.

For the LWP camera the largest non-linearities occur at the extreme
wavelengths (1900, 3300 \AA), where the flux is largely overestimated.
Except for these bands, linearity is within $\pm$ 5\% for spectra with
exposure levels from 40\% to 150\%.  In the saturated region of the most
exposed spectrum the flux is overestimated by 10\%.  Excluding the
saturated region, the bands which show the best linearity characteristics
(within $\pm$ 3\%) are those centered at 2800 and 3000 \AA.

The LWR camera shows the largest non-linearities at the longest wavelengths
of the underexposed spectra.  Linearity remains within $\pm$ 5\% for
exposure levels above 60\%.  The most linear bands are those centered at
2500, 2900 and 3100 \AA.  In the saturated part of the 200\%\ spectrum, the
flux is underestimated by approximately 10\%.  However, the flux is correct
in the 170\% spectrum, which is also saturated.

\section{Summary}
	\label{sec:summ}

Within the framework of the development of the ESA \ines\ Data Distribution
System foe the \iue\ Final Archive, \iue\ Low Dispersion spectra have been
re-extracted from the bi-dimensional SILO files with a new extraction
software. \ines\ implements a number of major modifications with respect to
the \swet\ extraction applied to the early version of the \iue\ Final
Archive.  The improvements in \ines\ deal with the noise model, the optimal
extraction method, the homogenization of the wavelength scale and the
flagging of the absolute calibrated extracted spectra.

\begin{itemize}

\item{} The noise models for the different cameras have been re-derived
to correct anomalies at  high and  low exposure levels in those
used in  \swet. The  noise model results in a considerably more
realistic estimate of the actual extraction errors in the \iue\ spectra.

\item{} The algorithms to compute the camera background and the
extraction profile are more consistent with the nature of the \iue\
detectors and result in a significantly improved data quality. 

\item{} Weak extended or miscentered spectra are 
more adequately handled. The fluxes of strong emission lines in weak
continuum spectra are more reliable and consistent in the \ines\
extraction.

\item{} The handling and propagation of quality flags to the final
extracted spectra has been improved. This implies a larger number of pixels
flagged, but also a more correct information for the user of potential
problems in the data.

\item{} A major improvement has been reached in the removal of the solar
contamination in LWP images after 1992.

\item{} In order to facilitate an easier use of the data, all the spectra
of a given range (short and long) have been resampled to a common wavelength
scale.

\end{itemize}

As a general rule, \ines\ data are similar or superior to \swet.  Although
the \ines\ spectra may at times give somewhat lower signal-to-noise ratio
than those obtained through \swet\ (e.g. when boxcar extraction is required
to maintain data validity), the \ines\ extraction results in a higher
reliability of the IUEFA data, allowing direct intercomparison of all low
resolution spectra, through an adequate treatment of errors, flags and
warning messages in the image header.

\begin{acknowledgements}

We would like to acknowledge the support of all VILSPA staff, which
collaborated actively to the development of the \ines\ system.

\end{acknowledgements}

\end{document}